\documentclass[manuscript]{aastex}
\tighten
\eqsecnum
\received{}
\accepted{}
\journalid{}{}
\articleid{}{}
\shortauthors{ }
\shorttitle{Pleiades Lithium} 
\begin{document}

\title{\ion{Li}{1} and \ion{K}{1} Scatter in Cool Pleiades Dwarfs\altaffilmark{1}}

\author{Jeremy R. King\altaffilmark{2}}

\author{Simon C. Schuler\altaffilmark{3}}

\author{L. M. Hobbs\altaffilmark{4}}

\author{Marc H. Pinsonneault\altaffilmark{5}}

\altaffiltext{1}{Based on observations obtained with the High Resolution Spectrograph on the
Hobby-Eberyly Telescope, which is operated by McDonald Observatory on behalf of the University
of Texas at Austin, Pennsylvania State University, Stanford University, the Ludwig-Maximillians-
Universitaet, Munich, and the George-August-Universitaet, Goettingen.  Public Access time was 
available on the Hobby-Eberly Telescope through an agreement with the National Science Foundation.}

\altaffiltext{2}{Department of Physics and Astronomy, Clemson University, 118 Kinard Lab, Clemson, SC  29634;
jking2@ces.clemson.edu}

\altaffiltext{3}{Leo Goldberg Fellow, Kitt Peak National Observatory, National Optical Astronomy Observatory, 
P.O. Box 26732, Tucson, AZ  85726-6732; sschuler@noao.edu.  NOAO is operated by the Association of Universities for Research in
Astronomy, Inc., under cooperative agreement with the National Science Foundation.}

\altaffiltext{4}{University of Chicago, Yerkes Observatory, 373 West Geneva Street, Williams Bay, WI  53191;
hobbs@yerkes.uchicago.edu}

\altaffiltext{5}{Department of Astronomy, Ohio State University, 140 West 18th Avenue, Columbus, OH  43210;
pinsono@astronomy.ohio-state.edu}

\begin{abstract}

We utilize high-resolution (R${\sim}60,000$), high S/N (${\sim}100$) spectroscopy of 17 cool Pleiades 
dwarfs to examine the confounding star-to-star scatter in the ${\lambda}6707$ \ion{Li}{1} 
line strengths in this young cluster.  Our Pleiads, selected for their small projected rotational
velocity and modest chromospheric emission, evince substantial scatter in the linestrengths of ${\lambda}6707$ 
\ion{Li}{1} feature that is absent in the $\lambda$7699 \ion{K}{1} resonance line.  The \ion{Li}{1} scatter is 
not correlated with that in the high-excitation ${\lambda}7774$ \ion{O}{1} feature, and the magnitude of the 
former is greater than the latter despite the larger temperature sensitivity of the \ion{O}{1} feature.  These 
results suggest that systematic errors in linestrength measurements due to blending, color (or color-based 
$T_{\rm eff}$) errors, or line formation effects related to an overlying chromosphere are not the principal 
source of \ion{Li}{1} scatter in our stars.  There do exist analytic spot models that can produce, via 
line formation effects, the observed Li scatter without introducing scatter in the \ion{K}{1} line strengths 
or the color-magnitude diagram.  However, these models predict factor of ${\ge}3$ differences in abundances 
derived from the subordinate ${\lambda}6104$ and resonance ${\lambda}6707$ \ion{Li}{1} features; we find no 
difference in the abundances determined from these two features.  These analytic spot models also predict CN 
line strengths significantly larger than we observe in our spectra.  The simplest explanation of the Li, K, 
CN, and photometric data is that there must be a real abundance component to the Pleiades Li dispersion.  We 
suggest that this real abundance component is the manifestation of relic differences in erstwhile pre-main-sequence 
Li burning caused by effects of surface activity on stellar structure. We discuss observational predictions 
of these effects, which may be related to other anomalous stellar phenomena.  

\end{abstract}


\section{Introduction}

Dramatic differences in the Li abundances of main sequence stars in open clusters stand in stark contrast to
the greater uniformity that is the general rule for many other elements.  The complexity of the observed pattern
of stellar Li depletion was recognized early \citet{GHH65,WHC65} and can be traced to the fragility of the
species.  Lithium is destroyed by proton capture at relatively low stellar interior temperatures
(of order 2.6 million K for typical densities); these conditions are achieved for most low mass stars
during the pre-main sequence (pre-MS) phase, which yields a predicted mass-dependent depletion pattern (e.g.,
Iben 1965).  Physical processes neglected in standard stellar models can also induce lithium depletion
\citep{WS65}.   Open cluster studies have revealed two other generic features of Li abundance
patterns: {\ }the existence of a dispersion in abundance at fixed mass, composition, and age and the
existence of main sequence depletion even in stars with convection zones too shallow to be able to
burn lithium (for reviews see Pinsonneault 1997 and Jeffries 2006).

The specific case of star-to-star Li dispersion in the Pleiades, and other young clusters such as IC 2602
\citep{Rand01b} and ${\alpha}$ Per \citep{Bala96}, has been extremely challenging to understand 
from both theoretical and observational perspectives.  A substantial dispersion in the equivalent width of 
the 6707 {\AA} \ion{Li}{1} resonance line for cool Pleiads was reported by \citet{DJ83} and confirmed by the 
much larger data set of \citet{Sod93}.   Subsequent observations of the weak 6104 {\AA} subordinate \ion{Li}{1} 
feature by \citet{FJS} yielded a consistent result.  The star-to-star Li equivalent width variations, which 
are superposed on a strong $T_{\rm eff}$-dependent depletion pattern that is presumably governed by pre-MS 
Li destruction in the deep surface convection zones whose extent is determined uniquely by stellar mass for 
a given composition and age in standard stellar models, have been difficult to explain in the context of 
differing surface Li abundances.  Star-to-star Li abundance variations can develop in the context of rotational 
mixing \citep{Pin90,RD95}, and are certainly seen in older open clusters such as M67 \citep{PRP,JFS99}.  
However, little such mixing is expected to have occurred at the Pleiades age (${\sim}100$ Myr).  Furthermore, 
the highest rate of mixing is expected in rapid rotators, while the rapid rotators tend to populate the upper 
envelope of the Li equivalent width distribution in the Pleiades.

It thus remains important to establish whether the star-to-star range in Pleiades' Li equivalent widths
reflect a real star-to-star range in Li surface abundances before strong conclusions can be drawn about
the underlying physical cause of the Li line strength dispersion.  While \citet{Sod93} suggest the Pleiades
Li dispersion reflects real abundance differences, they note the difficulty for this explanation that
comes from the scatter they find in the Pleiades' ${\lambda}7699$ \ion{K}{1} resonance feature.  \citet{J99}
confirm this \ion{K}{1} scatter, and note the concomittant danger in concluding that the Pleiades Li
dispersion stems from real abundance variations--a caveat echoed by \citet{Stu97} and \citet{Carl94}, who
note the similarity in the details of the Li and K resonance line formation.   \citet{KS04} catalog a rich
variety of Li-K and alkali-activity correlations in young clusters, seen in the Pleiades by \citet{KKP}, 
associating some 90\% of the variance in Li and K in M34 (200-250 Myr) and IC 2391 (50 Myr) with the spread 
in chromospheric emission (though this emission is likely a proxy for surface inhomogeneities in the case of 
M34).  Because rapid rotators in the Pleiades evince apparent Li overabundances \citep{Sod93,GL94}, chromospheric 
emission may simply serve as a proxy for rotation in this case.  In their study of the 100 Myr old NGC 2451 A 
and B, \citet{Marg02} find that significant Li scatter, reminiscent of that in the Pleiades, is spuriously produced
as an artifact of deriving Li abundances from equivalent widths; they suggest that accounting (via spectral
synthesis) for blending features in rotationally broadened spectra might eliminate the Pleiades Li scatter.

Here, we seek to address the important question of a real Li dispersion component in cool Pleiads by utilizing
new spectroscopy of higher resolution and/or S/N than in most previous studies to examine Li and K spreads
in slow projected rotators that demonstrate modest chromosperic emission relative to other cluster members.
We conclude that there is evidence for a real dispersion in abundance at fixed effective temperature, and (in
the final section) advance the idea that variations in pre-main sequence depletion stemming from 
differences in stellar physical parameters arising from surface inhomogeneities, rather than variations in 
the rate of mixing, may be implicated.

\section{Observational Data and Analysis}

We selected 17 Pleiads having a range of $T_{\rm eff}$ believed to evince significant Li dispersion 
(Soderblom et al.~1993; King, Krishnamurthi \& Pinsonneault 2000, hereafter KKP) and having 
$v{\ }sin{\ }i{\le}12$ km s$^{-1}$.  All objects are radial velocity members and have at least one 
proper-motion study indicating cluster membership.  Fourteen of the objects show no radial velocity or 
photometric evidence of binarity \citep{Merm92}.  \ion{H}{2} 571 and 2406 are 
single-lined binaries, but were included because their mass function determination and photometric 
decomposition suggest companion $V$-band contamination at only the ${\le}1$\% level.  \ion{H}{2} 298 is 
the lower mass 6-7 arcsec distant visual binary companion to \ion{H}{2} 299.  Our sample is listed in 
Table 1, which gives information on S/N, luminosity, color, membership, and rotational velocity.  Figure 1 
shows our objects in the \ion{Ca}{2} near-IR triplet- and H$\alpha$-based chromospheric emission versus 
$V_o$ planes defined by the Pleiades sample from KKP.  Our projected slow rotators exhibit well-below maximal 
chromospheric emission values at a given luminosity. 
\marginpar{Tab.~1}
\marginpar{Fig.~1}

We used the Hobby-Eberly Telescope 9.2-m telescope and its High Resolution Spectrograph (HRS) to obtain 
spectroscopy of our targets
on numerous nights from August 2002 to October 2003.  Wavelength coverage from 5095 to 8860 {\AA} was 
achieved over the HRS 2-CCD mosaic.  The S/N in the \ion{Li}{1} ${\lambda}6707$ region is 80-160 per pixel.
The 0.5 arcsec slit width yielded a nominal resolving power of $R{\sim}60,000$.  Standard reductions were
carried out with the {\sf IRAF} package to accomplish bias removal, scattered light subtraction, 
flat-fielding, order extraction, and wavelength calibration.  We do not conduct spectrum synthesis of 
the ${\lambda}6707$ \ion{Li}{1} and ${\lambda}7699$ \ion{K}{1} lines since quantitative deblending is not 
required for these features in our spectra and absolute abundances are not the principal topic of interest.  
Instead, we focus on equivalent width differences of these lines for any subset of the program stars having 
a narrow range of color.  

Equivalent widths of the ${\lambda}6707$ \ion{Li}{1} and ${\lambda}7699$ \ion{K}{1} resonance lines were 
measured with Guassian and Voigt profile fitting routines using the {\sf SPECTRE} package \citep{FS}.  
The principal potential danger in this approach is the mild blending some 0.4{\AA} to the blue of the 
\ion{Li}{1} feature.  However, its presence is of little importance here: the small projected rotational 
velocities mean that any residual contamination in the empirically {\it deblended\/} Li equivalent widths 
given in Table 1 is limited to a few m{\AA} (the size of random uncertainties in the line measurement), 
and the strength of the blending \ion{Fe}{1} (augmented by CN in cooler dwarfs) feature proves to be nearly 
invariant among Pleiads of given intrinsic color.  More 
importantly, even the full blend contribution is fractionally small, achieving strengths of 15-20 m{\AA} 
only in our coolest stars whose Li equivalent widths are an order of magnitude larger.  A conservative upper 
limit on equivalent width uncertainties including continuum placement is ${\le}10$ m{\AA}.  

For the purpose of full disclosure, we note that two spectra of \ion{H}{2} 152 taken 
14 October 2003 UT are clearly at least double-lined.  Two other spectra (08 December 
2002 UT) are single-lined.  The flux levels and gross spectral appearance do not
clearly establish a (queue) target misidentification for the 2003 spectra.  However, this 
seems the simplest explanation inasmuch as several previous independent high-resolution 
spectroscopic
studies (e.g., Soderblom et al.~1993; Wilden et al.~2002; Boesgaard, Armengaud \& 
King 2003) have not noted an SB2 classification.  While the BY Dra classification by 
\citet{Kholo89} is not inconsistent with binarity, it appears to be based on the modest 
photometric variation attributed to 4.1d variations from spot pattern migration by 
\citet{Mag87}.  We proceed here with the 2002 spectra and the assumption that the spectrum 
is single-lined.  

\section{Results and Discussion}

Our results are presented in Figure 2, containing purely observational planes showing 
equivalent widths versus (deredenned) $(B-V)$ color.  Several conclusions can be 
reached:\\ First, our new data nearly remove the star-to-star scatter in the 
${\lambda}7699$ \ion{K}{1} equivalent widths seen in \citet{Sod93} for our stars.  While 
measurement uncertainty is therefore important in the \citet{Sod93} results, this can not 
be the source of most of the star-to-star Li dispersion in their data since this scatter 
(up to a factor of 2 in equivalent width for $0.72{\le}(B-V)_0{\le}0.84$) persists in 
our own higher-quality data.  Second, the significant Li scatter in our projected slow 
rotators indicates that abundance errors due to blending features \citep{Marg02} do not 
provide the primary explanation of the Pleiades Li scatter.  
\marginpar{Fig.~2}

Finally, the vast extent of scatter in \ion{Li}{1} compared to \ion{K}{1} suggests that 
simple color or $T_{\rm eff}$ errors or the influence of an overlying chromosphere are 
not a significant source of star-to-star Li scatter in our Pleiads.  Color or temperature 
errors would lead to a similarly large dispersion in the similarly temperature-dependent 
\ion{K}{1} lines that is absent in our data.   Since the details of ${\lambda}6707$ 
\ion{Li}{1} and ${\lambda}7699$ line formation in cool dwarfs are similar 
\citep{Hou95,Stu97}, the dramatic difference between the Li and K in our Pleiads rules 
out differences in the global properties of overlying chromospheres as a sole or dominant 
source of the Li scatter.  Indeed, our sample selection was made to mitigate such 
effects:{\ \ }as seen in Figure 1, our stars evince modest scatter in chromospheric 
emission compared to a more representative cluster sample.  This is not to say that 
such differences are unimportant contributors to Li scatter in other samples \citep{KS04}.
Indeed, even here the \ion{K}{1} equivalent width of our reddest Pleiad (\ion{H}{2} 298), 
which also possesses the largest chromospheric emission index in our sample, is anomalously
large compared to a modest linear extrapolation of the \ion{K}{1}-(B-V) relation that 
can be seen to be linear as red as $(B-V)_0{\sim}1.4$ in Figure 1 of \citet{Rand01} for 
the 35 Myr IC 2602 cluster.  

\subsection{Photospheric Inhomogeneities}

It is expected that surface magnetic activity (spots and plages) could alter \ion{Li}{1} line 
strengths.  Spatially resolved solar observations show variations by factors of ${\sim}2$ and tens 
in the \ion{Li}{1} equivalent width in spot and plage regions, respectively \citep{G84}. 
\citet{Pat93} have found \ion{Li}{1} line strength variations in weak-lined T Tau stars that are 
not correlated with chromospheric emission variations, but are consistent in size with those expected
from the simple spot/plage model of \citet{G84}.  \citet{JBD94} find rotational modulation of the 
${\lambda}6707$ (and/or nearby blended features) in the young Local Association K6 dwarf BD$+$22 4409.  

One might think that the stark difference between the factor of ${\sim}2$ scatter in ${\lambda}6707$ 
\ion{Li}{1} line strengths for $0.7{\le}(B-V)_0{\le}0.8$ versus the absence of scatter in ${\lambda}7699$ 
\ion{K}{1} over this same range is a powerful argument against spots as a source of Li scatter in our stars.  
Surprisingly, this is not necessarily the case.  The influence of various analytic spot models on the 
location of Pleiads in the K,Li line strength versus color plane can be seen in Figures 3 and 6 of 
\citet{Barr01}.  The models with the two lowest photosphere-spot temperature contrasts (their models 2a 
and 2b) and 80\% spot coverage are able to displace stars redward and to higher \ion{Li}{1} equivalent 
width into the $0.7{\le}(B-V)_0{\le}0.8$ color range in such a way as to introduce a near factor of 2 
scatter in Li line strength, but move stars nearly parallel to the intrinsic \ion{K}{1} line strength 
versus color locus.  I.e., the analytic spot models {\it can\/}, in fact, produce the \ion{Li}{1} 
scatter we observe while {\it not} introducing substantial scatter in the K line strengths.  

As seen in Figure 2 of \citet{Barr01}, these models also move the Pleiads roughly parallel to the main-sequence 
in the H-R diagram such that significant photometric scatter, ${\Delta}V{\ge}0.1$ mag, is not introduced 
either; such scatter is not evinced by our stars (Figure 3).  Barrado y Navascues et al.'s (2001) larger 
photosphere-spot 
temperature contrast models (models 2c and 2d) can reproduce the observed scatter in Li line strengths 
with spot coverage fractions of ${\sim}60\%$; however, their Figure 2 shows that such stars would be 
subluminous by ${\Delta}V{\ge}1$ mag in the color-magnitude diagram.  The photometric data in Figure 3 clearly excludes
these large temperature contrast spot models.   The large photosphere-spot temperature contrast models only 
modestly alter the star's color (Figure 2 of Barrado et al. 2001).  As a result, a similar degree of scatter 
is introduced into the Li and K linestrengths (Figures 3 and 8 of Barrado et al. 2001)--a prediction also 
excluded by our spectroscopic data.  
\marginpar{Fig.~3}

In sum, the marked difference between \ion{K}{1} and \ion{Li}{1} scatter can not exclude a spot origin
for the Li scatter in our Pleiads.  While the indistinguishable spread of Li in the $(V-I)$- and $V$-Li
EW planes (Figure 4) compared to that in the $(B-V)$-based plane of Figure 2 does not betray the presence 
of spots, two spectroscopic signatures do robustly exclude the analytic spot models 2a and 2b of 
\citet{Barr01} as a source of the Li scatter.  The first is the difference in Li abundances derived
from the $\lambda$6707 \ion{Li}{1} resonance line and the weaker blended 1.85 eV $\lambda$6103.6 
\ion{Li}{1} subordinate feature(s) previously investigated in the Pleiades by \citet{FJS}.   We selected several 
interesting sets of Pleaids of similar color but with significantly different $\lambda$6707 \ion{Li}{1} line 
strengths (Table 2).  Effective temperatures and microturbulent velocities are adapted from 
the spectroscopic parameters determined by \citet{FJS}.  Following these authors, we assumed log $g=4.5$ for all 
stars since the derived Li abundances are insensitive to the assumed gravity.  We determined Li abundances from 
the ${\lambda}6707$ resonance line from our measured equivalent widths via Dr.~A.~Steinhauer's {\sf LIFIND} 
program discussed in \cite{KS05}; these are given in column 6 of Table 2.    
\marginpar{Fig.~4}
\marginpar{Tab.~2}

Abundances from the $\lambda$6104 \ion{Li}{1} features were determined via spectral synthesis carried out 
with an updated version of the LTE analysis MOOG \cite{Sneden73} and Kurucz model 
atmospheres\footnote{http://kurucz.cfa.harvard.edu/grids.html}.   The $\lambda$6104 region linelist was 
formed utilizing atomic data from the Vienna Atomic Line Database \citep{Kupka2000} and Kurucz line 
data\footnote{http://kurucz.cfa.harvard.edu/linelists.html}, and CN data from \citet{DP63} .  The linelist 
was calibrated by adjusting some oscillator strengths in order to produce solar syntheses matching the Kurucz 
solar flux atlas \citep{Kurucz2005} using input abundances from \citet{AG89} except
for CNO; solar values of log $N$(C)$=8.39$, log $N$(N)$=7.78$, and log $N$(O)$=8.68$ were adopted from
\citet{2005ASPC} and \citet{2001ApJ}.  The Pleiades syntheses assumed input scaled solar abundances 
with [m/H]$=-0.06$ based on the results of \citet{FJS} 

Sample ${\lambda}6104$ \ion{Li}{1} spectra and syntheses are shown in Figure 5.  Abundances are given 
in the penultimate column of Table 2, whose final column lists the ${\lambda}6104$- and ${\lambda}6707$-based
Li abundance differences.  There are two notable features in the abundance comparisons.  First, the 
significant Li scatter persists when measured from the ${\lambda}6104$ feature.  Second, we find no
significant difference between the resonance line- and subordinate line-based abundances.  The mean
difference (6104$-$6707) is $+0.01$ dex with small scatter (${\pm}0.08$ dex, s.d.).  
\marginpar{Fig.~5}

We find that the \citet{Barr01} analytic spot models reproducing the observed Li dispersion of our sample in 
the range $0.7{\le}(B-V)_0{\le}0.8$, lead to significantly larger predicted 6104$-$6707 Li differences.  Syntheses 
of the ${\lambda}6104$ and ${\lambda}6707$ \ion{Li}{1} regions--the latter using the linelist from \citet{K97} 
updated with VALD atomic data and recalibrated to the solar flux spectrum-- were performed using the photosphere 
and spot parameters of models 2a and 2b with 80\% spot coverage from \citet{Barr01} assuming log $g=4.5$, 
${\xi}=2.0$ km s$^{-1}$, [m/H]$=-0.06$, and an input abundance of log $N$(Li)$=2.80$.  The photospheric and 
spot spectra were weighted by their respective Planck functions and coverage fractions before adding and 
renormalizing.  

The resulting ${\lambda}6104$ and ${\lambda}6707$ photosphere$+$spot spectra were analyzed as 
observed data via spectrum synthesis computed for $T_{\rm eff}=5225$ (the value of the spotted models consistent 
with flux conservation), log $g=4.5$, ${\xi}=2.0$, and [m/H]$=-0.06$.  The ${\lambda}$6707 Li abundances deduced 
from the spotted synthetic spectrum are log $N$(Li)$=2.90$ and 3.02 for spot models 2a and 2b, respectively.  The 
corresponding ${\lambda}$6104 Li abundances are 3.32 and 3.62.  The Li differences (6104$-$6707 {\AA}) deduced from 
the two \ion{Li}{1} lines-- 0.42 and 0.6 dex for spot models 2a and 2b--are significantly larger than the zero 
difference exhibited by our stars. 

The second spectroscopic signature that is inconsistent with the analytic spot models reproducing
the observed Li spread in our stars is the strength of CN lines in both the ${\lambda}6104$ and ${\lambda}6707$ 
region.  The CN line depths in the $T_{\rm eff}=5225$ spotted synthetic spectra are some 3-5 times deeper 
than observed in our actual object spectra.  Figure 6 illustrates a related important point: while 
the warmer photosphere ($T_{\rm phot}=5655$ K; 20\% coverage) and cool spots ($T_{\rm spot}=4870$ K; 80\% 
coverage) in model 2a of \cite{Barr01} conspire to yield a spotted star of $T_{\rm eff}{\sim}5225$, the
spectrum of such a model is not equivalent to that of an unspotted model with $T_{\rm eff}=5225$.  Figure 6
indicates the CN features are significantly stronger in the spotted model.    
\marginpar{Fig.~6}

\subsection{Over-Excitation and -Ionization Effects}

\citet{Sch2004} find that our cool Pleiads' O abundances derived from the high-excitation \ion{O}{1} 
${\lambda}7774$ 
triplet in our spectra show a dramatic increase with declining $T_{\rm eff}$ and significant star-to-star 
scatter.  \citet{Sch2006} show that photospheric hot spots provide a plausible explanation for this behavior.  In
order to examine whether the Li and O scatter are related, the ${\lambda}6707$-based Li and ${\lambda}7774$-based
O abundances were fit as functions of $T_{\rm eff}$ using low order polynomials.  To ensure consistency for this 
purpose, Li abundances were determined with {\sf LIFIND} using the stellar parameters of \citet{Sch2004}; these
were updated for \ion{H}{2} 298 using our $(B-V)_0$ value, adjusting the \citet{Sch2004} O abundance 
(to [O/H]$=+0.80$) in the process.  The resulting abundance residuals (observed minus fitted), shown in Figure 7, 
are not correlated.  Moreover, the rms dispersion of the observed Li abundances (${\sim}0.22$ dex) is twice that 
of the O abundances (${\sim}0.12$ dex) despite the fact that the temperature sensitivity of the (logarithmic) O 
abundances is 2-3 times larger than that for Li.  We conclude that the dispersion in our Pleiads' Li is not 
associated with that in \ion{O}{1}.  
\marginpar{Fig.~7}

\section{Conclusions, a Proposed Explanation, and Future Work}

\subsection{Summary and Key Conclusions} 

We have utilized high-resolution and -S/N spectroscopy to measure ${\lambda}6707$ \ion{Li}{1} and 
${\lambda}7699$ \ion{K}{1} line strengths and ${\lambda}6707$- and ${\lambda}6104$-based Li abundances in 
17 slowly (projected) rotating cool Pleiades dwarfs.  A significant factor of ${\sim}2$ dispersion is seen
in the ${\lambda}6707$ \ion{Li}{1} line strengths over the $0.72{\le}(B-V)_0{\le}0.82$ color range; this large 
Li scatter is also inferred from ${\lambda}6104$ subordinate line-based abundances.  The 
scatter in our selected sample eliminates line blending due to rapid projected rotation as a source of Li 
dispersion in our Pleiads.  In contrast to previous studies, our high-resolution and S/N data evince no substantial 
scatter in the ${\lambda}7699$ \ion{K}{1} line strengths of our particular Pleiad sample.  This stark distinction 
relative to the Li linestrengths excludes simple color or $T_{\rm eff}$ errors or line formation effects due to an 
overlying chromosphere as the source of Li scatter in our stars.   

The difference in the dispersions of the \ion{K}{1} and \ion{Li}{1} line strengths does not, however, exclude
spots as a source of the Li scatter; in particular, the analytic spot models 2a and 2b of \citet{Barr01} can
produce the Li scatter we observe without leading to significant scatter in \ion{K}{1} line strengths or in
the color magnitude diagram.  The equivalence of the ${\lambda}6707$- and ${\lambda}6104$-based Li abundances 
in our stars does, however, exclude these spot models, which predict a factor of ${\ge}3$ difference in 
the resonance and subordinate line-based Li abundances.  These spot models also would lead to CN linestrengths
significantly larger than observed in our spectra.   

The simplest explanation of the \ion{Li}{1} dispersion in our Pleiads is scatter due to real abundance differences.  
There are numerous candidate mechanism(s) to explain these pre- or near-ZAMS abundance differences:{\ \ }chemical 
inhomogenieties, magnetic field differences, variable mass accretion, and individual rotational 
histories \citep{V98,GL94}.  While the effects of chromospheres on \ion{Li}{1} line formation and the effects of 
rotation on measurements of the \ion{Li}{1} line strength espoused by \citet{KS04} and \citet{Marg02} may contribute 
to an (illusory) Li dispersion, our results suggest that there must also be a real component to the Pleiades Li 
dispersion--at least over the range $0.7{\le}(B-V)_0{\le}0.8$ seen in Figure 2.  

The size of the real component is difficult to determine directly; however, future work can deduce 
the size and mass range of real scatter since sources of possible illusory dispersion are amenable to 
observation.  Systematic effects of rotation on Li abundance {\it measurement} of rapid rotators or from 
low spectral resolution (neither characterize our data) can be mitigated by determining abundances from 
spectrum syntheses with suitable accounting of macroscopic broadening.  The influence of chromospheric 
emission and surface magnetic activity (spots and plages) on \ion{Li}{1} (and \ion{K}{1}) lines can be 
measured by searching for correlated temporal variations in alkali line strength and chromospheric 
emission and photometric indices via a simultaneous spectroscopic and photometric monitoring program of 
the Pleiades \citep{J99}.  Our Li equivalent widths and those of \citet{Sod93} for 5 stars (\ion{H}{2} 152, 
263, 2126, 2311, and 2366) differ at the ${\ge}2{\sigma}$ level, providing impetus for such a program.  Additional 
constraints may be gleaned from observations of $^9$Be in our cool Pleiads, as well as in similarly cool Hyades 
stars, whose degree of Li scatter is unknown due to vanishingly weak Li line strengths \citep{Sod95}.  

\subsection{A Proposed Differential Pre-MS Li Burning Mechanism}

We believe that the most likely theoretical explanation of a real star-to-star Li dispersion in the Pleiades 
is a range in pre-MS lithium depletion.  Such a range would naturally emerge if the radii of protostars of 
the same mass, composition, and age during the epoch of pre-MS lithium depletion were somehow different.  To 
explain the empirical trend, one would also require the more rapidly rotating and active stars to have 
been larger.  There is now emerging evidence that activity impacts the radii of both main sequence and pre-MS 
stars in precisely this sense.  \citet{TR02} found that the radii of the near-twin stars in YY Gem were far 
too large to be consistent with the predictions of standard stellar theory.  Subsequent data confirmed this
result, and \citet{MRJ08} found a correlation between activity and the radius excess in low mass stars.  
\citet{Berg06} also found this to be a relatively common phenomenon in their measurements of the radii of 
field M stars, although it is more challenging to perform a rigorous test in the absence of direct mass 
constraints in such systems.  \citet{AP04}, in a study of cataclysmic variable systems, found that excess 
activity could induce radius changes of the proper order of magnitude (see also Chabrier, Gallardo, \& 
Baraffe 2007).  

The physical mechanism linking activity-related radii differences with those in Li depletion is that spots 
inhibit convective energy transport, requiring the star to carry the flux through a smaller effective volume.  
The star therefore requires a larger radius, corresponding to an effective reduction in the mixing length or 
efficiency of energy transport.  This in turn reduces pre-MS lithium burning because the central temperature 
and pressure of a star is decreased if the radius is increased.  We performed a simple numerical test of this 
phenomenon.  A 10\% difference in radius (of order seen in active eclipsing binaries) during standard pre-MS 
convective burning for a 0.9 M$_{\odot}$ solar abundance model led to a factor of 2.5 increase in the remaining 
surface lithium abundance at an age of 100 Myr relative to an inactive model with the same mass but a smaller radius.
How the spot properties of single pre-MS stars (the precursors of our Pleiad sample) compare to those of active 
binaries remains an open question.  Some meaningful context, however, might be provided by chromospheric 
emission fluxes.  We note that the log $R'$(HK) indices for active solar-type binaries widely range from $-4.4$ 
to $-3.3$ at a given color or Rossby number (Figure 4 of Montes et al.~1996), while the corresponding Pleiades 
(pseudo-)indices (which must be transformed from H$\alpha$ and \ion{Ca}{2} infrared triplet data) widely range 
from -4.5 to -3.5 (Figure 3 of King et al.~2003); i.e., active binaries show wide chromospheric emission at 
the same levels and with similar wide dispersion as present-day Pleiads. 

When/if spot filling factors are reduced subsequent to pre-MS Li depletion, stars would converge to similar 
radii, but would retain a fossil record of their differences in earlier stages.  If more spotted stars on the 
main sequence retained their lower effective temperatures, one would also shift more massive (and less depleted) 
stars to the same effective temperature as less massive (and more depleted) bare stars, further amplifying 
the apparent Li dispersion.  There is evidence that activity can differentially impact protostellar radii.  
\citet{SMV07} found a reversal in the predicted $T_{\rm eff}$-radius relationship between a more active primary 
and a less active secondary of a brown dwarf binary system in the young (1 Myr) Orion Nebula Cluster. 

Our suggested explanation of the Pleiades Li dispersion has several attractive features.  It explains why 
the dispersion is large amongst cooler stars, which experience significant pre-MS depletion, compared to hotter
stars, and the propensity for more rapidly rotating (and perhaps more spotted) cool dwarfs to exhibit larger 
Li line strengths.  It also can qualitatively explain the apparent narrowing of dispersion in cool main sequence 
stars in older open clusters as follows.  Beyond the Pleiades age, more heavily spotted and rapdily rotating 
stars deplete more Li than barer and more slowly rotating stars due to rotationally-induced main-sequence mixing; 
as spottedness declines with increasing age and stellar radii concomitantly relax, the former stars also become 
hotter relative to the latter stars.  The combined effect would be a reduction in the star-to-star dispersion 
in cool cluster Li abundances that is controlled by the timescales of stellar angular momentum loss and 
spot coverage dimunition.  Because the former timescale could, in principle, be longer than the latter, the 
passage of additional time would then witness the resurgence of substantial Li dispersion due to rotationally-induced 
main-sequence mixing acting in stars whose radii and temperatures are no longer scattered due to surface 
inhomogeneities.  This is not inconsistent with the observation that the ratio of rapid and slow rotators in 
the ${\sim}100$ Myr Pleiades is the inverse of that of Li-rich and -poor solar mass stars in the 4.6 Gyr 
cluster M 67 \citep{JFS99}.  Our suggested explanation, which recognizes the importance of surface 
inhomogeneities on stellar radii and possible differences in time evolution of rotationally-induced mixing and 
spot coverage, dovetails with the important observational picture painted by \citet{JFS99}:  a significant Li 
dispersion in cool dwarfs in very young clusters such as the Pleiades that markedly declines in intermediate 
age clusters such as M34 and the Hyades, and then reappears in older clusters such as M 67.  We caution, though,
that the dwarf Li abundances in M 67 are available only for stars of $T_{\rm eff}{\ge}5500-5600$ K; abundances 
in lower mass M67 dwarfs akin to our cooler Pleiads have not yet been determined.  It should also be noted that 
a variety of observational studies (e.g., Piau, Randich \& Palla 2003; Randich et al.~2007, Jeffries
et al.~2002) have suggested other mechanisms besides or in addition to rotationally-induced mixing as the source of 
main-sequence Li depletion and explaining the resulting main-sequence Li-$T_{\rm eff}$ morphologies manifested by 
open cluster data. 

\subsection{Related Stellar Phenomena and Future Work}
 
It is also tempting to speculate that the starspot-radius mechanism might be connected with other puzzling 
stellar phenomena.  The alleged existence of a gap (or gaps) in the number distribution of Pop I main sequence stars 
\citet{Men56,BV70,RC2000} around $(B-V){\sim}0.3$ could be accomodated if this color corresponds to the onset 
of surface spot formation driving stars to lower $T_{\rm eff}$ values compared to bluer barer stars.  The 
accompanying increase in stellar radii might give rise to the surprising number of luminosity class 
II-IV stars that reside in the main-sequence region of the Hipparcos-based H-R diagram \citep{NY98}.  The 
precipitous rise in apparent [\ion{Fe}{2}/H] abundances and more gradual but nevertheless surprising rise 
in apparent [\ion{Fe}{1}/H] abundance in the Hyades with declining $T_{\rm eff}$ \citep{Y94}, the rise in 
high-excitation permited [\ion{O}{1}/H] abundances with declining $T_{\rm eff}$ in the Hyades and the even
steeper rise in the younger Pleiades \citep{Sch2006}, the rise in forbidden line-based [\ion{O}{1}/H] abundances
with declining $T_{\rm eff}$ in the Hyades \citep{Sch2006b}, the rise in [\ion{Si}{1}/H] abundances with 
declining $T_{\rm eff}$ in M 34 \citep{Sch2003} are at least qualitatively consistent with a $T_{\rm eff}$-dependent
disparity between spot-adjusted radii and assumed standard stellar radii.  Such a disparity would lower 
log $g$ with declining $T_{\rm eff}$; the sensitivity of the features noted above is such that lowering 
log $g$ increases their line strengths in cool dwarfs. 

Future observational work is needed to confirm these speculative connections; quantifying or parameterizing 
spot coverage and properties rather than chromospheric emission indices {\it per se} that are used as measures 
of ``activity'' will be of particular importance.  Such work includes: continued exploration of the reality of 
main-sequence gaps in open clusters, their evolution with age, and the spot properties of stars adjacent to 
these gaps; understanding the nature of luminosity class II-IV stars, in particular spot properties and physical 
radii, near the main-sequence region of the H-R diagram; the spot properties of stars as a function of 
$T_{\rm eff}$ in the Pleiades, M 34, and Hyades open clusters, and their association with the abundance anomalies 
(especially star-to-star variations in permitted line [\ion{O}{1}/H] values in cool Pleiads and Hyads) listed above. 

A possible signature of the mechanism we propose here is the Li/Be ratio in stars sufficiently cool and old
to suffer main-sequence mixing of $^9$Be, but not so cool as to have suffered pre-MS convective burning of $^9$Be.  
Such rotationally-induced main-sequence mixing would deplete more Be (and Li) in stars with larger initial
rotational velocities; however, if such stars were more heavily spotted, they suffered reduced pre-MS convective 
burning of Li; depending on the balance of spot-induced retarded pre-MS Li depletion and rotation-induced enhanced 
MS Li depletion, this effect could create a population of higher-Li/lower-Be stars whose number depends upon the 
initial distribution of rotation, its time evolution, and age.  Identifying such stars using extant disk field 
abundance data is difficult given differences 
in stellar structure due to metallicity differences, unknown relative initial Li and Be abundances, and a dispersion 
in age.  Such a search is best carried out in cool dwarfs of open clusters of at least intermediate (e.g., Hyades) 
age.  For the purpose of constraining the origin of Li scatter in the Pleiades, a monitoring program to identify short- 
(days) and long- (years) term variations in \ion{Li}{1} linestrengths and spot properties is critical.  If our 
proposed origin of the Pleiades Li dispersion is correct, we expect to see variations in Li line strength that 
are consistent only with changes in photospheric structure/parameters and the details of spot-included radiative transfer 
alone; once these variations are accounted for, a significant dispersion in Li line strength should remain as a relic 
of spot-induced differences in pre-MS Li burning.   While observationally challenging, these future observational 
programs (accompanied by detailed modeling of pre-MS Li depletion in stars of various spotted conditions) will be 
required to fully understand the magnitude, mass distribution, and thus the ultimate origin of real star-to-star Li 
dispersion in the Pleiades. 
  
\acknowledgments
JRK and SCS gratefully acknowledge support for this work from NSF awards AST-0086576 and AST-0239518, 
a generous grant from the Charles Curry Foundation to Clemson University, and a graduate scholarship award 
from the South Carolina Space Grant Consortium.

%%Fig 1
\begin{figure}
\plotone{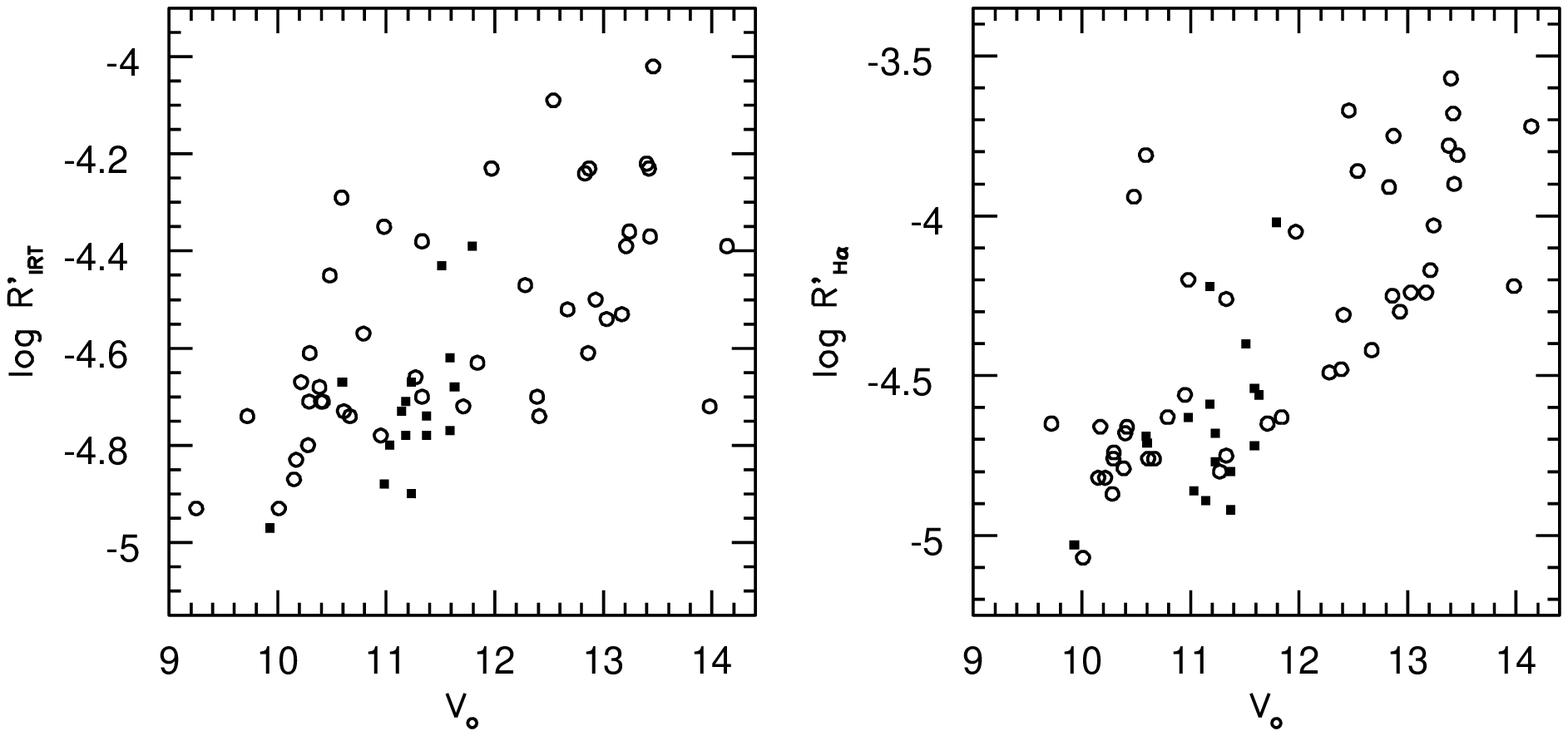}
\caption[]{Our Pleiades sample (filled squares; Table 1) is plotted with other Pleiads from the study of 
KKP in the chromospheric emission (left: \ion{Ca}{2} infrared triplet; right: H$\alpha$) versus $V_o$ planes.  
The chromospheric fluxes are taken from Soderblom et al.~(1993).} 
\end{figure}

%%Fig 2
\begin{figure}
\plotone{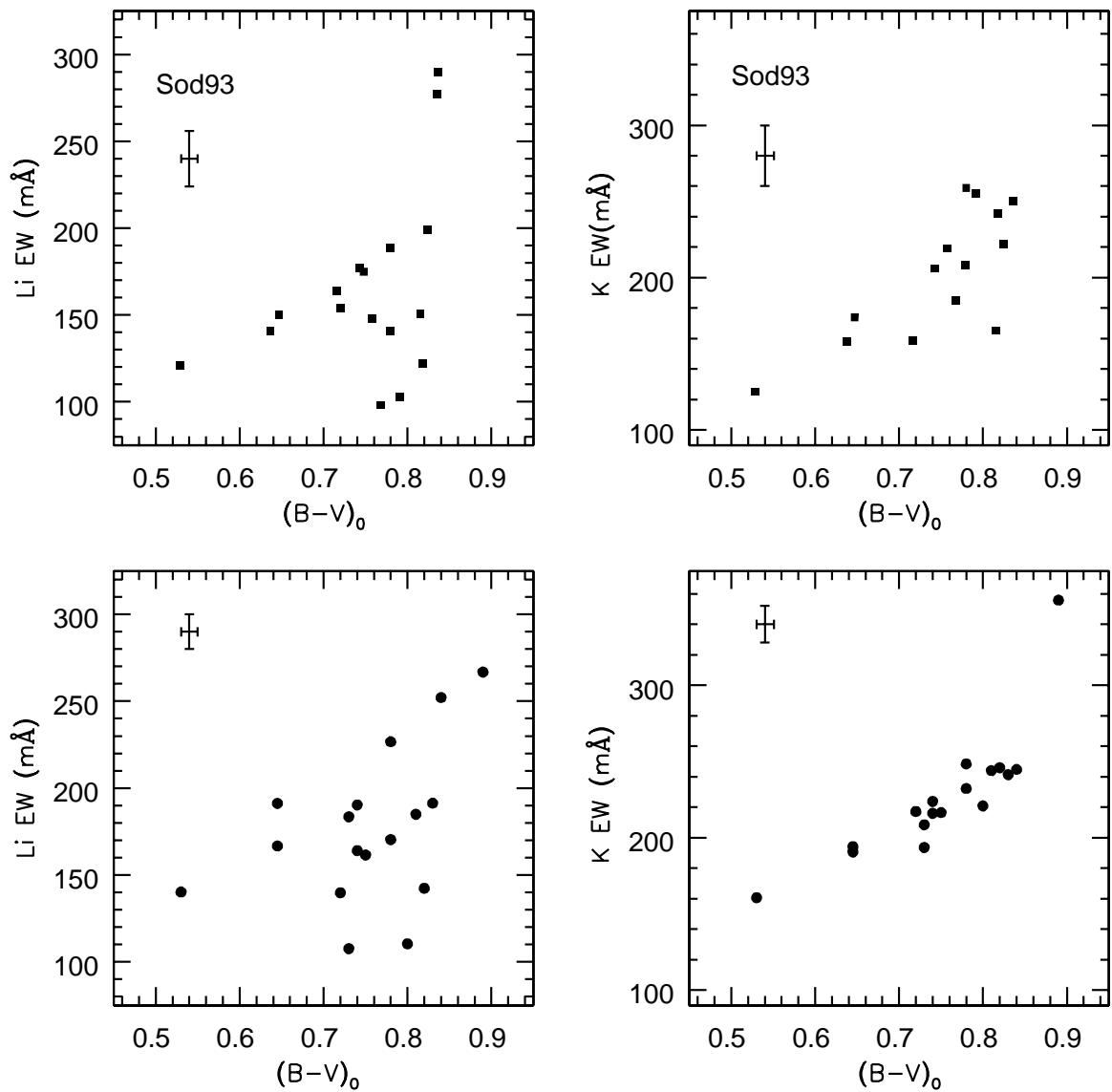}
\caption[]{The top panels show ${\lambda}6707$ \ion{Li}{1} (left) and ${\lambda}7699$ \ion{K}{1} (right) 
equivalent widths from \citet{Sod93} versus dereddened $(B-V)$ color.  The bottom panels show the 
(empirically deblended in the case of Li) equivalent widths measured from our HET/HRS spectra. Representative
error bars are shown in the upper left of each panel.} 
\end{figure}

%%Fig 3
\begin{figure}
\plotone{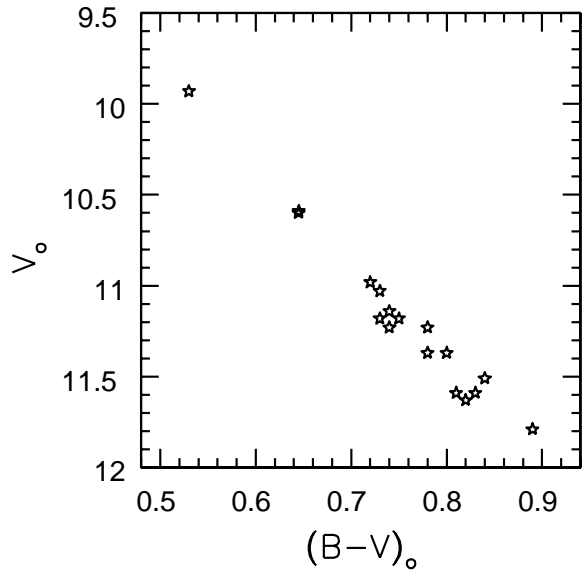}
\caption[]{A color-magnitude diagram of our Pleiades stars. The ${\ge}1$ mag scatter in $V$ implied by large 
photosphere-spot temperature contrast spot models of \citet{Barr01} that might reproduce the Li line strength 
dispersion in Figure 2 is not present; any modest photometric scatter is not inconsistent with Barrado et 
al.'s smaller temperature contrast spot models, however.}
\end{figure}

%%Fig 4
\begin{figure}
\plotone{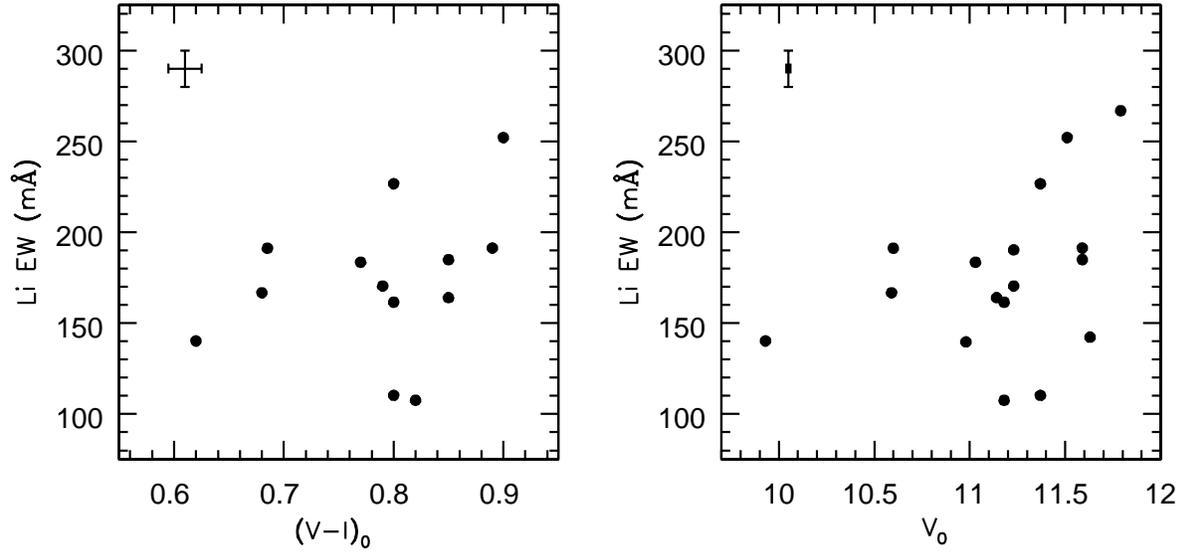}
\caption[]{The ${\lambda}6707$ \ion{Li}{1} equivalent widths of our Pleiades sample versus $(V-I)$ color (left) 
and $V$ magnitude (right).  Typical uncertainties are shown in the upper left of both panels.} 
\end{figure}

%%Fig 5
\begin{figure}
\plotone{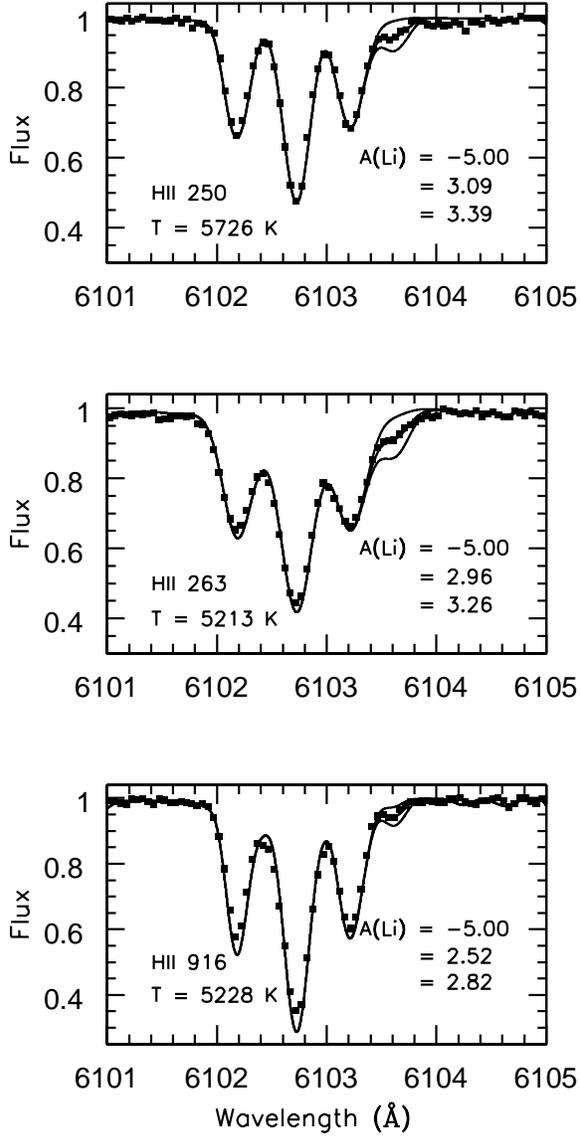}
\caption[]{Observed (points) and synthetic (lines) spectra of the ${\lambda}6103.6$ \ion{Li}{1} region in 
\ion{H}{2} 250, 263, and 916 (top, middle, and bottom panels).  The syntheses are shown for 3 input Li 
abundances for each star:  no Li, the best fit Li abundance, and a Li abundance 0.3 dex larger than the best 
fit value.} 
\end{figure}

%%Fig 6
\begin{figure}
\plotone{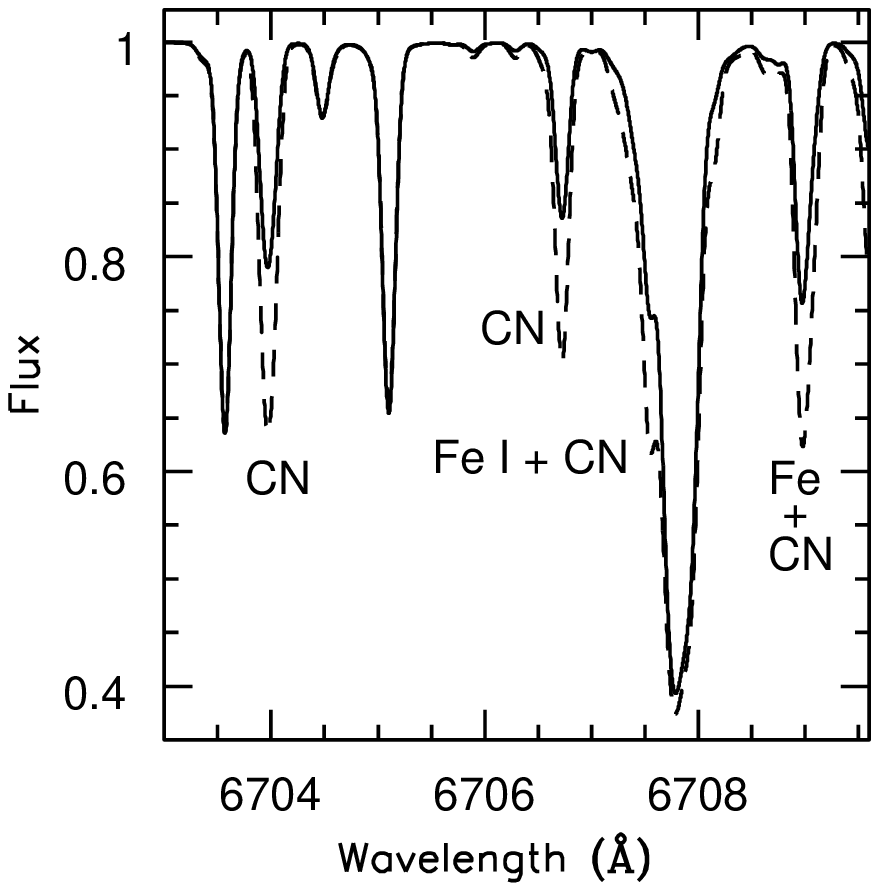}
\caption[]{(Dashed line) A synthetic spectrum of the ${\lambda}6707.8$ \ion{Li}{1} region for the \citet{Barr01} 
analytic spot model 2a with $T_{\rm phot}=5655$ K (20\% coverage) and $T_{\rm spot}=4870$ K (80\% coverage), 
yielding a spotted model having $T_{\rm eff}{\sim}5225$ K.  (Solid line) A synthetic spectrum for 
$T_{\rm eff}=5225$ K.  All syntheses employ a 0.11 {\AA} FWHM gaussian smoothing and assume identical input 
abundances (solar) except for Li, for which log $N$(Li)$=2.8$ was utilized.  The CN features are significantly 
stronger in the spotted model.}
\end{figure}

%%Fig 7 
\begin{figure}
\plotone{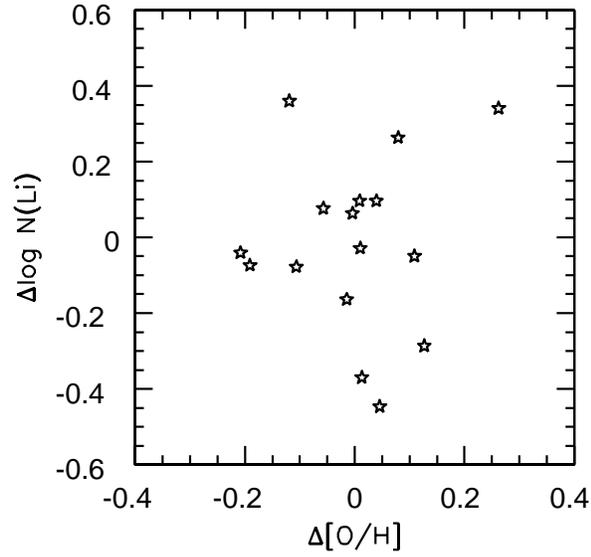}
\caption[]{Residuals of Pleiades \ion{Li}{1} abundances derived from our line strengths about a low-order $T_{\rm eff}$-dependent 
polynomial versus the residuals in ${\lambda}7774$-based \ion{O}{1} abundances derived from the same spectra \citep{Sch2006}.}
\end{figure}   

%%TABLE 1 
%\documentclass{aastex}
%\begin{document}
\begin{deluxetable}{lrrrrrrrrr}
\tablecolumns{9}
\tablewidth{0pc}
\tablenum{1}
\tablecaption{HET Pleiades Sample} 
\tablehead{
\colhead{Star} & \colhead{S/N} & \colhead{$V_o$} & \colhead{$(B-V)_o$} & \colhead{$(V-I)_o$} & \colhead{Mem. Probs.\tablenotemark{a}} & 
\colhead{RV\tablenotemark{b}} & \colhead{$v$ sin $i$\tablenotemark{c}} & \colhead{EW(\ion{K}{1})} & \colhead{EW(\ion{Li}{1})} \\
\colhead{\ion{H}{2}} & \colhead{\@6707} & \colhead{ } & \colhead{ } & \colhead{ } & \colhead{BFJ,SRSG,DH} & \colhead{km s$^{-1}$} & 
\colhead{km s$^{-1}$} &  \colhead{m\AA} & \colhead{m\AA}  
}
\startdata
0152 & 96  & 10.60 & 0.645 & 0.685 & --,0.98,--     & $5.2{\pm}0.2$ & 11.5      & 194.0 &  191.2 \\
0193 & 100 & 11.18 & 0.75  & 0.80  & 0.97,0.95,--   & $7.6{\pm}0.2$ & 6.6       & 216.4 &  161.5 \\
0250 & 97  & 10.59 & 0.645 & 0.68  & --,0.98,0.57   & $4.7{\pm}0.3$ & 6.4       & 190.6 &  166.7 \\
0263 & 133 & 11.51 & 0.84  & 0.90  & 0.98,0.98.0.17 & $3.1{\pm}0.9$ & 7.8       & 244.7 &  252.1 \\
0298\tablenotemark{d} & 80 & 11.79 & 0.89  & *     & --,--,--       & $4.8{\pm}0.2$ & 6.5  & 355.9 & 266.8 \\ 
0571 & 126 & 11.14 & 0.74  & 0.85  & 0.97,0.97,0.3  & $5.7{\pm}0.1$ & 7.2       & 223.8 & 164.0 \\ 
0746 & 140 & 11.18 & 0.73  & 0.82  & 0.99,0.98,--   & $6.5{\pm}0.4$ & 4.9       & 208.4 & 107.6 \\
0916 & 109 & 11.59 & 0.83  & 0.89  & 0.99,0.98,0.90 & $4.2{\pm}0.7$ & 6.2       & 241.3 & 191.3 \\
1593 & 128 & 11.03 & 0.73  & 0.77  & 0.99,0.04,0.01 & $6.9{\pm}0.1$ & 2.4       & 193.5 & 183.5 \\ 
2126 & 119 & 11.59 & 0.81  & 0.85  & 0.98,0.00,0.05 & $5.5{\pm}0.2$ & ${\le}6$  & 244.0 & 184.9 \\ 
2284 & 105 & 11.23 & 0.74  & *     & 0.99,0.97,--   & $6.3{\pm}1.1$ & 3.6       & 216.0 & 190.3 \\
2311 & 114 & 11.23 & 0.78  & 0.79  & 0.96,0.99,0.38 & $5.4{\pm}0.2$ & 6.4       & 232.3 & 170.4 \\
2366 & 81  & 11.37 & 0.78  & 0.80  & --,0.98,--     & $6.2{\pm}0.2$ & ${\le}5$  & 248.4 & 226.7 \\ 
2406 & 150 & 10.98 & 0.72  & *     & 0.98,0.98,--   & $6.0{\pm}0.1$ & 8.9       & 217.1 & 139.7 \\ 
2462 & 110 & 11.37 & 0.80  & 0.80  & 0.99,0.96,--   & $6.4{\pm}0.2$ & 5.2       & 220.8 & 110.3 \\ 
2880 & 84  & 11.63 & 0.82  & *     & 0.98,0.97,--   & $5.4{\pm}0.2$ & 6.2       & 245.9 & 142.3 \\ 
3179 & 158 &  9.93 & 0.53  & 0.62  & --,0.98,--     & $6.1{\pm}0.2$ & ${\le}6$  & 160.5 & 140.2 \\ 
\tablenotetext{a}{Proper motion-based membership probabilities from \citet{BFJ}, \citet{SRSG}, and \citet{DH}} 
\tablenotetext{b}{Mean radial velocities from WEBDA database: http://obswww.unige.ch/webda}
\tablenotetext{c}{Rotational velocities (or conservative upper limits) from \citet{Q98}} 
\tablenotetext{d}{$V$ and $(B-V)$ from the photometric decomposition of \citet{Merm92}} 
\enddata
\end{deluxetable}                 

%%TABLE 2
\begin{deluxetable}{lrcccccc}
\tablecolumns{8}
\tablewidth{0pc}
\tablenum{2}
\tablecaption{${\lambda}6707$ \ion{Li}{1} versus ${\lambda}6104$ \ion{Li}{1}}
\tablehead{
\colhead{Star} & \colhead{$(B-V)_o$} & \colhead{$T_{\rm eff}$} & \colhead{$\xi$} & \colhead{EW(6707)} & \colhead{A(Li)} & \colhead{A(Li)} & \colhead{$\Delta$A(Li)} \\
\colhead{\ion{H}{2}} & & K & km s$^{-1}$ & m{\AA} & ${\lambda}6707$ & ${\lambda}6104$ & ${\lambda}6104-{\lambda}6707$ 
}
\startdata
250  & 0.645 & 5726 & 2.00 & 166.7 & 3.03 & 3.09 & 0.06 \\
     & & & & & & & \\  
746  & 0.73  & 5383 & 2.10 & 107.6 & 2.31 & 2.33 & 0.02 \\
1593 & 0.73  & 5383 & 2.10 & 183.5 & 2.75 & 2.73 & -0.02 \\
2284 & 0.74  & 5450 & 2.88 & 190.3 & 2.87 & 2.90 & 0.03 \\
     & & & & & & & \\
2311 & 0.78  & 5350 & 2.30 & 170.4 & 2.64 & 2.77 & 0.13 \\
2366 & 0.78  & 5350 & 2.30 & 226.7 & 2.98 & 2.91 & -0.07 \\  
     & & & & & & & \\
2462 & 0.80  & 5283 & 2.15 & 110.3 & 2.21 & 2.37 & 0.16 \\
2880 & 0.82  & 5250 & 2.15 & 142.3 & 2.36 & 2.30 & -0.06 \\
2126 & 0.81  & 5267 & 2.15 & 184.9 & 2.61 & 2.55 & -0.06 \\
     & & & & & & & \\
916  & 0.83 & 5228 & 2.15 & 191.3 & 2.60 & 2.52 & -0.08 \\
263  & 0.84 & 5213 & 2.13 & 252.1 & 2.97 & 2.96 & -0.01 \\
\enddata
\end{deluxetable}

\end{document}